\documentclass[apj]{emulateapj}
\usepackage{natbib,graphicx,epstopdf,url}
\shortauthors{Plotkin et al.}

\begin{document}

\title{An Environmental Study of the Ultraluminous X-ray Source Population in Early-type Galaxies}
\shorttitle{ULXs in Early-type Galaxies}

\author{
Richard.~M.~Plotkin\altaffilmark{1},
Elena~Gallo\altaffilmark{1},
Brendan P.~Miller\altaffilmark{1,2},
Vivienne F. Baldassare\altaffilmark{1},
Tommaso Treu\altaffilmark{3},
and Jong-Hak Woo\altaffilmark{4}
}

\altaffiltext{1}{Department of Astronomy, University of Michigan, 500 Church St., Ann Arbor, MI 48103, USA; rplotkin@umich.edu}
\altaffiltext{2}{Department of Physics \& Astronomy, Macalester College, 1600 Grand Avenue, Saint Paul, MN 55105, USA}
\altaffiltext{3}{Physics Department, University of California, Santa Barbara, CA 93106, USA}
\altaffiltext{4}{Astronomy Program, Department of Physics and Astronomy, Seoul National University, Seoul, Republic of Korea}
\newcommand{\nh}{$N_{\rm H}$}    
\newcommand{\lledd}{$L_X/L_{\rm Edd}$}  
\newcommand{\ledd}{L_{\rm Edd}}   
\newcommand{\ergs}{{\rm erg~s}^{-1}}
\newcommand{\rgas}{r_{\rm gas}}       

\newcommand{\mdot}{$\dot{m}$}
\newcommand{\Mdot}{$\dot{M}$}
\newcommand{\msun}{M_{\odot}}
\newcommand{\mstar}{M_{\star}}

\newcommand{\nar}{NewAR}

\begin{abstract}
Ultraluminous X-ray sources (ULXs) are some of the brightest phenomena found outside of a galaxy's nucleus, and their explanation typically invokes accretion of material onto a black hole.   Here, we perform the largest population study to date of ULXs in early-type galaxies, focusing on whether a galaxy's large scale environment can affect its ULX content.   Using the AMUSE survey, which includes homogeneous X-ray coverage of 100 elliptical galaxies in the Virgo cluster and a similar number of elliptical galaxies in the field (spanning stellar masses of $10^8-10^{12}~\msun$), we identify $37.9\pm10.1$ ULXs in Virgo and  $28.1\pm8.7$ ULXs in the field.  Across both samples, we constrain the number of ULXs per unit stellar mass, i.e., the ULX specific frequency, to be $0.062 \pm 0.013$ ULXs per 10$^{10}~\msun$ (or about 1 ULX per $1.6\times10^{11}~\msun$ of galaxy stellar mass).  We find that the number of ULXs, the specific frequency of ULXs, and the average ULX spectral properties are all similar in both cluster and field environments.  Contrary to late-type galaxies, we do not see any trend between specific ULX frequency and host galaxy stellar mass, and we show that dwarf ellipticals host  fewer ULXs than later-type dwarf galaxies at a statistically meaningful level.    Our results are consistent with ULXs in early-type galaxies probing the luminous tail of the low-mass X-ray binary population, and are briefly discussed in context of the influence of gravitational interactions on the long-term evolution of a galaxy's (older) stellar population.

\end{abstract}


\section{Introduction} 

  Ultraluminous X-ray sources (ULXs) are extragalactic, non-nuclear point sources, typically defined by  X-ray luminosities $L_X>10^{39}~{\rm erg~s}^{-1}$ (0.3-10~keV).  ULXs radiate well above the  Eddington luminosity\footnote{The Eddington luminosity is $\ledd = 1.26\times10^{38}\left(M/\msun\right)~\ergs$ for ionized Hydrogen, assuming that the emission is isotropic, where $M$ is the mass of the accreting object in Solar masses.} 
 for a neutron star  and are  almost certainly accreting black holes, but their exact nature is  still highly debated.    A handful of the very brightest ULXs ($L_X \gtrsim 10^{41}~{\rm erg~s}^{-1}$) are viable intermediate mass black hole candidates (IMBHS), potentially representing the missing link between  stellar mass and supermassive black holes (e.g., HLX-1 in ESO243-49 is one of the best candidates so far; \citealt{farrell09}).  Other possible explanations include black hole X-ray binaries (BHXBs) with unusually massive black holes ($M\sim 20-100~\msun$) formed by the direct collapse of a metal-poor star \citep[e.g.,][]{belczynski10}, or BHXBs with a relativistically beamed jet pointed toward the observer  (i.e., stellar mass analogs to blazars; although see, e.g., \citealt{davis04, kaaret04, kaaret09} for reasons why the ``micro-blazar'' scenario is unlikely).
  
 While the above possibilities are very intriguing, the vast majority of ULXs do not  require  such exotic explanations.  Allowing for anisotropic emission and super-Eddington accretion up to 10 times larger than the Eddington limit \citep{begelman02}, the normal BHXB population can  account for ULXs with X-ray luminosities up to  $\sim10^{40}~{\rm erg~s}^{-1}$.  Most ULXs thus simply   represent the high-luminosity tail of  a galaxy's normal ($\lesssim 20\msun$) BHXB population \citep{feng11}.    Ignoring the more exotic types of ULXs with $L_X \gtrsim 10^{40}~{\rm erg~s}^{-1}$, ULXs are predominantly composed  of two BHXB populations --- young high mass X-ray binaries (HMXBs)  in galaxies actively forming stars, and transient low mass X-ray binaries (LMXBs) generally found in ellipticals \citep{king02}.  Population studies on ULXs therefore provide an economic means to study the BHXB content of galaxies, which  can place constraints on a galaxy's evolution and star formation history. 
   
Given typical sensitivities of X-ray facilities like \textit{Chandra} and \textit{XMM-Newton}, systematic searches for ULXs are  limited to galaxies within a few tens of Mpc \citep[e.g.,][]{swartz11}.    Since the Local Volume contains mostly late-type  galaxies, overall ULX   number counts and X-ray luminosities are   generally consistent with expectations from HMXB luminosity functions, with ULXs more likely to be found in galaxies with higher global star formation rates (SFRs; e.g., \citealt{grimm03}).   This trend with SFR implies that the large-scale environment may influence the properties of ULXs, especially considering that the most  luminous ULXs are generally found in interacting galaxies \citep{feng11}.  Furthermore, as  shown by \citet{swartz08}, the number of ULXs per unit stellar mass ($\mstar$) of the host galaxy  (which we refer to as $S_{\rm ulx}$, the specific ULX frequency) is anti-correlated with stellar mass (for $\mstar>10^{8.5}~\msun$, below which not enough galaxies have yet been probed to  constrain the anti-correlation).   This anti-correlation surprisingly implies that ULXs reside relatively more  frequently  in dwarf galaxies, which could be due to late-type dwarfs having higher specific-SFRs \citep[e.g.,][]{brinchmann04}, and/or ULXs in dwarfs being longer-lived.  Regardless, the higher $S_{\rm ulx}$ is likely reflecting differences between the evolution of dwarf and giant (late-type) galaxies (e.g., dwarfs typically evolve more slowly with  lower metallicities and less merger-dependent histories; see \citealt{swartz08} and references therein for the relevance to ULXs).

   Only about one-third of known ULXs are found in early-type galaxies \citep{walton11, feng11}.   The lower number of ULXs in ellipticals appears to be more than an observational artifact due to surveying only the Local Volume, as the specific ULX frequency is also 10 times smaller in ellipticals  \citep{swartz11,walton11}.   \citet{irwin04} show that  ULXs  in early-type galaxies  also tend to be  less luminous than ULXs in late-type galaxies (with perhaps all ULXs in ellipticals having $L_X < 2 \times 10^{39}~{\rm erg s}^{-1}$).   In addition,  contrary to spirals, elliptical galaxies seem to show a flat trend  (or potentially a positive correlation) between specific ULX frequency and host stellar mass \citep{walton11}.     These observations are consistent with the ULX population in elliptical galaxies being associated exclusively with LMXBs, and ULXs in spirals being associated primarily with HMXBs (although at the lower-luminosity end, LMXBs can also contribute some fraction to a spiral galaxy's total ULX population; \citealt{colbert04}).  The number of ULXs in ellipticals should thus correlate with galaxy mass rather than SFR \citep{gilfanov04}.
         
Here, we  focus exclusively on the ULX population in early-type galaxies, which is currently not as well constrained as for late-types.  Besides improving their ULX number statistics (especially for galaxies with lower-stellar masses), our primary goal is to perform a focused study on whether there is a connection between ULXs and  large-scale environment.    Environmental effects strongly influence the evolution of galaxies (and their cold gas content) within clusters through interactions with the intracluster medium, with the cluster potential, and also via galaxy-galaxy interactions, through which star formation can be triggered and/or quenched (see e.g., \citealt{treu03}; \citealt{boselli06} for reviews on environmental processes).  Therefore, if ULXs reflect information about a galaxy's evolutionary history, we might  expect different ULX populations within early-type galaxies in cluster versus   isolated (field) environments.   For this  study, we use \textit{Chandra} X-ray observations of 195 early-type galaxies from the Active Galactic Nucleus (AGN) MUltiwavelength Survey of Early-Type Galaxies (AMUSE).  AMUSE is a unique resource because it contains a large  number of galaxies in both cluster and field environments across a wide range of stellar masses  ($10^{8}-10^{12}~\msun$).  We describe the AMUSE samples and X-ray data analysis in \S \ref{sec:obs}, we present results in \S\ref{sec:results}, and we discuss our results in \S \ref{sec:disc}.  Unless stated otherwise,  statistical errors are quoted at the 90\% confidence level.

 \section{The AMUSE Sample and X-ray Observations}
 \label{sec:obs}
 The AMUSE survey was designed  to use \textit{Chandra} to search for weak AGNs at the centers of early-type galaxies and is composed of two parts, the AMUSE-Virgo survey (targeting cluster environments) and the AMUSE-Field survey (targeting  more isolated environments).  The galaxies included in each subsample were optically selected and  therefore unbiased with respect to their X-ray properties.  Besides including a large number of galaxies spanning a wide range in $\mstar$ and in environment, a key property of the AMUSE survey is that it provides homogeneous X-ray coverage for each galaxy. The initial  AMUSE-Virgo survey  (ID 8900784, \textit{Chandra} Cycle 8, PI: Treu, 454 ks) targeted 100 spheroidal galaxies within the \textit{Hubble Space Telescope (HST)} Advanced Camera for Surveys (ACS) Virgo Cluster Survey \citep[VCS;][]{cote04} with \textit{Chandra ACIS-S} and \textit{Spitzer} MIPS \citep{gallo08, gallo10, leipski12}.    The AMUSE-Virgo \textit{Chandra} observations include 84 ``snapshots'' with exposure times $>$5.4~ks, which was supplemented by 16 deeper archival \textit{Chandra} observations.    After completion of AMUSE-Virgo, a complementary sample of early-type galaxies in non-cluster environments was targeted with a large \textit{Chandra} program, the AMUSE-Field survey (ID 11620915, \textit{Chandra} Cycle 11, PI: Gallo, 479 ks).  For AMUSE-Field, a total of 103  galaxies were optically selected from the HyperLeda\footnote{\url{http://leda.univ-lyon1.fr/}} catalog \citep{paturel03} for \textit{Chandra} follow-up (see \citealt{miller12a} for details).   The AMUSE-Field survey consists of $\sim$2--12~ks \textit{Chandra} ``snapshots'' for 61 galaxies from the Cycle 11 program, which are supplemented by archival \textit{Chandra} observations of 42 more early-type galaxies \citep{miller12a, miller12}.   Two-band \textit{HST/ACS} images were also taken for 28 AMUSE-Field galaxies (Baldassare et al.\ in prep).    The AMUSE-Virgo and AMUSE-Field X-ray data have ($3\sigma$)  sensitivity thresholds better than $3.75\times10^{38}~\ergs$ and $2.5 \times 10^{38}~\ergs$, respectively,  over the \textit{Chandra} bandpass, providing ample sensitivity to constrain an X-ray population with $L_X > 10^{39}~{\rm erg~s}^{-1}$.

\begin{figure}
\includegraphics[scale=0.85]{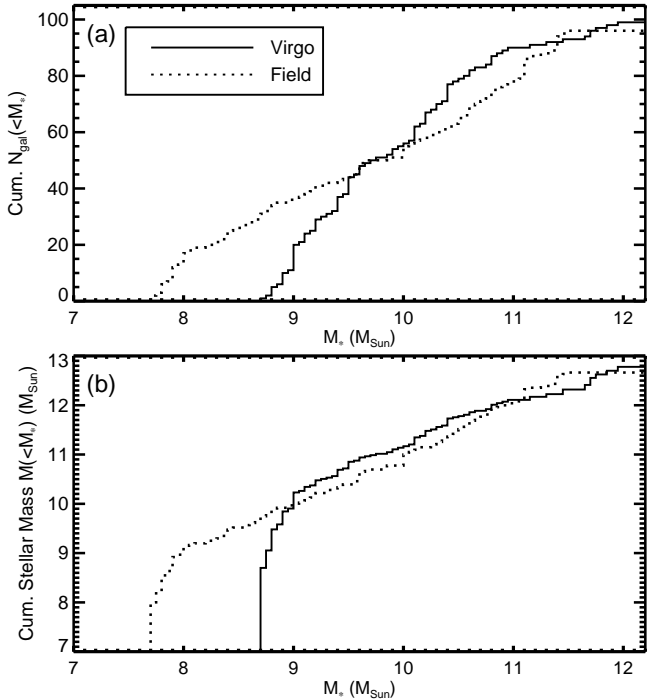}
\caption{Cumulative Number (a) and stellar mass (b)  distributions for galaxies in the Virgo (solid lines) and Field (dotted lines) samples. }
\label{fig:mass}
\end{figure}

A handful of the non-``snapshot''  \textit{Chandra} observations were read-out in subarray mode (VCC2095 = NGC4762 in the Virgo sample; NGC4036, NGC5322, and NGC5838 in the Field sample).  These observations are not considered here because they do not cover enough of the galaxy to search for non-nuclear X-ray sources.   Four additional Field galaxies are excluded (NGC3928, NGC3265, NGC0855, and ESO540-014) because their high-spatial resolution \textit{HST} images reveal complex morphologies.  It is now apparent that these galaxies are unlikely to be early-type, and they will be discussed in a future publication (Baldassare et al.\ in prep).  We thus consider 99 and 96 galaxies in the Virgo and Field samples, respectively.  Basic properties of these 195 galaxies are presented in Tables~\ref{tab:virgoprop}--\ref{tab:Fieldprop}, with stellar masses and distances taken from \citet{gallo10} and \citet{miller12a} for the Virgo and Field galaxies, respectively.\footnote{The stellar masses are derived from two-band optical photometry using the relations in \citet{bell03}.  \textit{HST/ACS} imaging in the F457W and F850LP filters, which roughly correspond to SDSS $g$ and $z$, are used when available. Otherwise, stellar masses are calculated from (in order of preference) $g-z$ model magnitude colors from the SDSS, or from $B-V$ colors taken from HyperLeda.  Distances for the Virgo galaxies are derived from surface brightness fluctuations \citep{mei07}, or set to 16.5~Mpc (the average distance to the Vigo Cluster) when such distances are unavailable.  For the Field galaxies, radial-velocity distances from HyperLeda are used, since redshift-independent distances are not available for most objects in either HyperLeda or NED.} 
The cumulative distributions of the number of galaxies in each sample and their stellar masses are shown in Figure~\ref{fig:mass}.  We also provide in Tables~\ref{tab:virgoprop}--\ref{tab:Fieldprop} the shape of each galaxy's 25~mag~arcsec$^{-2}$ isophote in the $B$-band, which we  refer to as the $D_{25}$ isophote, as taken from HyperLeda and corrected for extinction.  We  refer to the (angular) radius of the major axis of the $D_{25}$ isophote as $r_{25}$ throughout the text.

\subsection{X-ray Data Analysis}
\label{sec:data}
The \textit{Chandra} data reduction was performed with {\tt CIAO} version 4.5  \citep{fruscione06} and done in a similar fashion as described  by \citet{gallo08, gallo10} for AMUSE-Virgo, and by \citet{miller12a} for AMUSE-Field.  We refer the reader to those papers for details, and we only briefly describe the data reduction here.   All targets were placed on the back-illuminated S3 chip of the Advanced CCD Imaging Spectrometer (ACIS) detector \citep{garmire03}.   If a galaxy falls within the Sloan Digital Sky Survey (SDSS) footprint, we first improve the \textit{Chandra} astrometry by cross-matching X-ray point sources on the ACIS-S3 chip to the SDSS (the X-ray source lists are derived by running \textit{wavdetect} on the pipeline level 2 event files).  We require at least three matches to an SDSS optical counterpart (with optical magnitude $r<23$~mag), and we improve the astrometry and generate new aspect solution files for 70\% of the observations (all astrometry corrections are at the sub-pixel level, i.e., $<0\farcs5$).

Next, we reprocess each observation to generate new level 2 event files with the latest calibration applied, and we remove time intervals with anomalously high background rates ($>3\sigma$ above the mean level).  Then we create 0.3-7.0~keV X-ray images of the S3 chip from the reprocessed event files,  and we generate a list of point-sources in each image by running \textit{wavdetect} with scales of 1.0, 1.4, and 2.0 pixels using a 1.5~keV exposure map and a threshold significance of $10^{-6}$ (corresponding to approximately one false detection expected per chip).    Source counts (0.3--7~keV) are calculated for each wavdetect source from aperture photometry using circular apertures with radii corresponding to the 90\% encircled energy fraction at 1.5~keV (the encircled energy fraction is a function of each source's  off-axis position on the S3 chip).  The local background for each source is estimated from the median number of counts in four nearby (source-free) regions.   For 20 of the galaxies, there are significant amounts of hot diffuse gas near the central regions  that could contaminate the photometry.   Since the hot gas component contributes predominantly to the soft X-ray band, we use 2--7~keV hard-band images  to perform photometry for sources near the inner regions of these galaxies (adopting apertures with 90\% encircled energy fractions at 4.5~keV).  More details are provided in \S \ref{sec:gassims}, including an analysis to determine if the presence of hot gas can potentially bias our results.  In addition to the 20 galaxies with hot gas, we also use hard-band images to calculate net counts for sources in the inner 17$\arcsec$ of VCC1903 and NGC3379, and in the inner 32$\arcsec$ of NGC1052.  Although these three galaxies do not (obviously)  show significant diffuse emission, their inner regions are relatively crowded with point sources.  Since all three galaxies have relatively deep \textit{Chandra} exposures, using the hard-band images allows a cleaner estimate of the local background near each X-ray source.

Finally, given the observed number of source and background counts within each aperture (after applying a 90\% aperture correction), we assess which sources should be considered X-ray detections.  To determine the detection threshold on a source-by-source basis, we use the Bayesian formalism of \citet{kraft91} when the number of background  counts is less than 10, and we use Equation~9 from \citet{gehrels86} elsewhere.  We exclude all X-ray point sources below the detection threshold from further analysis.

\subsection{Identifying ULXs}
\label{sec:ulxcat}
To identify ULXs, we consider all X-ray point sources within each galaxy's $D_{25}$ isophote and more than 2$\arcsec$ from the galaxy's optical center (the latter constraint is to exclude potential AGNs).   There are a total of 705 and 1092 off-nuclear X-ray sources associated with galaxies in the Virgo and Field samples, respectively.  For each source, we use the measured 0.3-7~keV net count rates (2-7~keV for sources embedded in hot gas or in crowded fields) and version 4.6b of the Portable, Interactive Multi-Mission Simulator (PIMMS)\footnote{\url{http://heasarc.gsfc.nasa.gov/docs/software/tools/pimms.html}} to estimate 0.3-10~keV X-ray fluxes (corrected for Galactic absorption).   We assume an absorbed powerlaw with Galactic absorption (from the \citealt{dickey90} \ion{H}{1} maps)  and a photon index $\Gamma=1.8,$\footnote{The photon index is defined by $N(E)=N_0(E/E_0)^{-\Gamma}$, with $N(E)$ the number of photons at an energy $E$, $N_0$ the photon number normalization, and $E_0=1$~keV the reference energy.} 
   which is the average ULX photon index from \citet{swartz04}.   We then calculate luminosities for each point source based on the above fluxes and the distance to each galaxy.  We consider as ULXs all sources with unabsorbed 0.3-10~keV $L_X > 10^{39}~\ergs$,  retaining 55 and 50 ULX candidates in the Virgo and  the Field samples, respectively (before removing potential contaminants; see \S\ref{sec:bg}).   ULX number counts ($N_{\rm ulx}$) are provided in Tables~\ref{tab:virgoprop}--\ref{tab:Fieldprop} for each galaxy, and properties of each individual ULX candidate are given in Tables~\ref{tab:virgoulx}--\ref{tab:Fieldulx}.  
   
\subsubsection{Fraction of Each Galaxy Covered by the ACIS-S3 Chip}
\label{sec:frac}
For a small number of galaxies (16 in total), part of the outer regions of the $D_{25}$ isophote extends off the S3 chip (typically missing only $6\%$ or less of the galaxy's total projected area on the sky).   We expect only a few ULXs to populate the outer regions of galaxies (see Fig.~12 of \citealt{swartz04}), and the partial coverage should therefore not cause us to miss many ULXs.  Nevertheless, we  correct for any potential incompleteness by first calculating the fraction of each galaxy that is covered by the S3 chip ($f_{\rm area}$), determined by the fraction of illuminated pixels within the $D_{25}$ isophote compared to the total area of the $D_{25}$ isophote.    Then, we determine the  fraction of ULXs in each of these 16 galaxies that we expect to fall on the S3 chip ($f_{\rm ulx}$).  We assume that ULXs follow a surface density profile of the form $dN/dA \propto \exp^{-\left(16.67 r^\prime\right)^{0.63}}$, based on the empirical fit in \S3.2.4 of \citet{swartz04}.   $dN/dA$ is the number of ULXs per unit area, and $r^\prime=\left(r/r_{25}\right)$ is the dimensionless distance to an elliptical isophote with semi-major axis $r$ normalized to $r_{25}$.  We  calculate $f_{\rm ulx}$ as  the weighted fraction of pixels within the $D_{25}$ isophote covered by the S3 chip,  weighting each pixel by the above surface density profile.  The quantities $f_{\rm area}$ and $f_{\rm ulx}$ are provided in Tables~\ref{tab:virgoprop}--\ref{tab:Fieldprop} for each galaxy.   We  expect to recover  an average fraction of $\left<f_{\rm ulx}\right>=0.985$ ULXs among the 16 galaxies not fully covered by the S3 chip (with all 16 having $f_{ulx}>0.922$).  The ACIS-S3 chip covers 100\% of the other  179 AMUSE galaxies, so this source of incompleteness is very minimal when considering the entire sample.

\subsubsection{Contamination from the Cosmic X-ray Background}
\label{sec:bg}
 Here, we  assess the expected number of chance alignments of  foreground/background sources within each galaxy's $D_{25}$ isophote.  We use X-ray source counts from the resolved cosmic X-ray background study by \citet{moretti03}.  We first estimate the expected hard X-ray flux $S$ (2-10~keV) that would be observed from a $10^{39}~\ergs$ ULX (0.3-10~keV) in each galaxy using PIMMS, assuming the distance to each galaxy, $\Gamma=1.8$, and Galactic absorption.   Then, we use the \citet{moretti03} cumulative X-ray flux distribution (their Equation~2) to calculate $N(>$$S)$, the expected number of X-ray sources (per deg$^{2}$) with a hard X-ray flux larger than $S$.  Finally, $N(>$$S)$ is multiplied by the fractional area of the $D_{25}$ isophote covered by the ACIS-S3 chip for each galaxy ($f_{\rm area}$).    The expected number of background sources ($N_{\rm bg}$) is included in Tables~\ref{tab:virgoprop}-\ref{tab:Fieldprop}.  $N_{\rm bg}$ is negligible for the majority of galaxies, but the background contamination can be as high as 2--3 sources for the largest galaxies.    If $N_{\rm bg}>N_{\rm ulx}$ for a galaxy, then we set $N_{\rm bg}=N_{\rm ulx}$ to avoid negative net ULX counts.  We note that this constraint biases our net ULX counts to larger numbers, but it does not  affect our results qualitatively (see \S \ref{sec:numcounts}).    Across the entire Virgo and Field samples, we expect a total of 17.6 and 21.9 background contaminants, respectively.   We also  search the SIMBAD\footnote{The SIMBAD database (\url{http://simbad.u-strasbg.fr/simbad}) is operated at CDS, Strasbourg, France.} and NED\footnote{The NASA/IPAC Extragalactic Database (NED) is operated by the Jet Propulsion Laboratory, California Institute of Technology, under contract with the National Aeronautics and Space Administration.}  
databases to identify any ULX candidates that are already known to be background/foreground objects.   We identify four contaminants in the Virgo sample (three AGN and one star), and one AGN in the Field sample.    For completeness, these five sources are  included in the ULX count rate numbers ($N_{\rm ulx}$) in Tables~\ref{tab:virgoprop}--\ref{tab:Fieldprop}, and also included (and marked)  in the ULX catalogs (Tables~\ref{tab:virgoulx}--\ref{tab:Fieldulx}).  Ultimately, we correct for the background contamination statistically.

\subsection{Galaxies with Diffuse Gas: Assessing the Completeness Fraction}
\label{sec:gassims}
Some of the \textit{Chandra} images of the more massive galaxies in the AMUSE sample show a diffuse hot X-ray gas component  (these  tend to be  archival observations,  the majority of which have longer exposure times than the \textit{Chandra} snapshots).   The presence of hot gas could negatively affect our ULX study in two ways.  First, if not properly accounted for, the hot gas  could cause us to overestimate an X-ray point source's luminosity.  Second, it is possible that the hot gas could outshine a ULX that is embedded in the diffuse emission,   causing us to miss some ULXs and thereby underestimate ULX number counts (preferentially) for the most massive galaxies.   We address the first concern by performing photometry on hard-band images (2-7~keV)  for point sources near significant amounts of gas, as we expect the gas to contribute less than 5\% of the total flux at energies above 2~keV for most galaxies \citep[see][]{gallo08}.  To address the second concern, we use {\tt MARX}\footnote{\url{http://space.mit.edu/cxc/marx/}} version 5.0 to project ray-tracings of simulations of $10^{39}~\ergs$ point sources embedded in a  diffuse gas component onto the ACIS-S3 detector.  Details are described below.  

\begin{figure}
\includegraphics[scale=0.85]{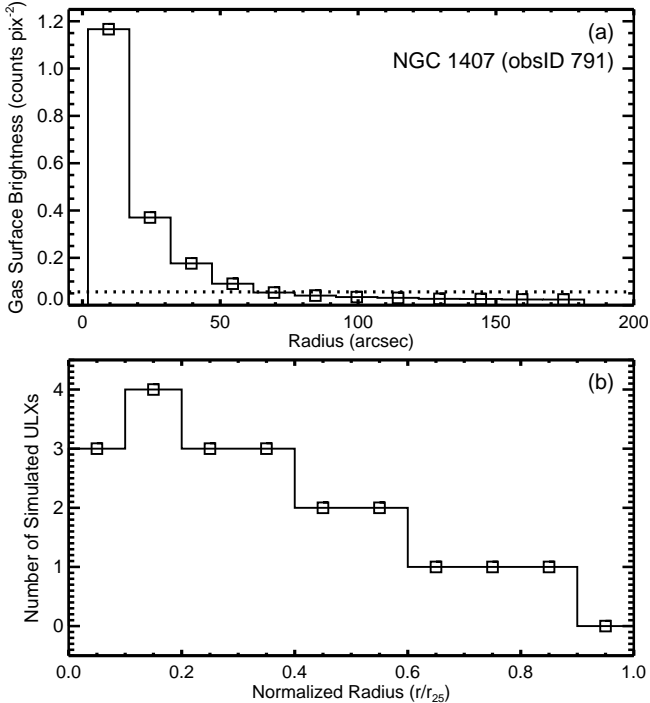}
\caption{(a) Sample radial profile for the extended gas emission in NGC1407 in the Field (uncertainties are smaller than the data symbols).  The dotted line shows the surface brightness of the background (see \S \ref{sec:gassims}).   For NGC 1407, we consider the gas to extend to $r_{\rm gas}=62\arcsec$  ($r_{\rm gas}/r_{25} = 0.33$).  (b) Adopted radial distribution for simulated ULXs.}
\label{fig:gasradprof}
\end{figure}

To determine which galaxies show significant gas emission, we first generate 0.3-7~keV band images of each galaxy masking out all point sources detected by {\tt wavdetect}.\footnote{We also mask out the X-ray jet for VCC1316 (M87) in Virgo.}  
  Then, we create a surface brightness radial profile, $\Sigma(r)$ (in counts per  deg$^2$), for each galaxy.  Galaxies with significant gas are easily identified by their smoothly declining radial profiles, while galaxies without gas have flat  profiles.    The radial profiles are calculated in radial bins of of width $\Delta r=15\arcsec$, and they extend from $r=2\arcsec$ to $r=r_{25}$.    We also estimate a background surface brightness level, $\Sigma_{bg}$,  from a circular aperture extending from 5-15$\arcsec$ past $r_{25}$.  We consider a galaxy to have significant gas at all radii where $\Sigma(r) > 2.5\Sigma_{bg}$, and we identify a gas boundary radius $\rgas$ as the angular distance of the outer edge of the annulus where $\Sigma(\rgas)=2.5\Sigma_{bg}$.    In total, there are 9 Virgo and 11 Field galaxies with significant gas, with $\rgas$ ranging from 1--6$\arcmin$ (or normalized to the $D_{25}$ isophote, $0.16 < \rgas/r_{25}< 0.77$).   A sample gas radial profile is shown in Figure~\ref{fig:gasradprof}a.

In order to simulate the hot gas component for these 20 galaxies, we first need a spectral model for the gas.  We  extract a gas spectrum for each galaxy from the above masked images with the {\tt CIAO} task {\tt specextract}, using a circular source extraction annulus with an inner radius=2$\arcsec$ and an outer radius set to $\rgas$ (and employing the same background region as described above).  Weighted background and source response matrix files (rmfs) and auxiliary response files (arfs) are created for each observation, and the spectra are fit using {\tt ISIS} version 1.6.2-10 \citep{houck00}, binning each spectrum to achieve a signal-to-noise $>$5 per bin.   The diffuse gas is fit with the Astrophysical Plasma Emission Code (APEC) thermal-emission model \citep{smith01}, i.e.,  {\tt vapec}, assuming Solar abundances \citep{anders89} and allowing the abundances of oxygen, neon, magnesium, silicon, and iron to vary as free parameters.  The plasma model  is calculated at the redshift of each galaxy, and we freeze the intervening column density to the Galactic value.   All galaxies' gas components can be adequately fit with  a single temperature plasma model, with gas temperatures ($kT_g$) ranging from 0.48-0.95 keV.  Only VCC1316 (M87) has substantially hotter gas, with $kT_g=1.60\pm0.02$~keV.

We next create a map of each galaxy's diffuse emission to use as a pixel-by-pixel spatial model for the {\tt MARX} simulations.  We start by generating a 0.3-7~keV band image of each galaxy extending out to a radius $\rgas$.  We use the {\tt CIAO} task {\tt dmfilth} to fill in regions of the image where {\tt wavdetect} identified a point source, interpolating the expected number of gas counts from a random sampling of  nearby pixels.\footnote{For VCC1316 (M87) in Virgo, we also fill in regions covered by the X-ray jet.}  
   From the above spatial maps and best-fit spectral models, we simulate each galaxy's diffuse gas emission with {\tt MARX},  centering the gas to the same location on the ACIS-S3 chip as in the real observations.  
   
   Next, we randomly simulate the positions of ULXs embedded inside each galaxy's gas component by assuming that ULXs follow the same surface density profile, $dN/dA$, as adopted in \S \ref{sec:frac}.  We normalize the  surface density profile to include 20 ULXs in each galaxy,\footnote{We intentionally overestimate the number of ULXs per galaxy in order to assess the statistical completeness of our ULX identification algorithm.}  
and then we numerically integrate $dN/dA$ over area to determine the expected number of ULXs within several radial bins (see Figure~\ref{fig:gasradprof}b).    We randomly draw pairs of right ascensions and declinations out to $r=\rgas$ (assuming random position angles), matching the number of ULXs within a given radial bin to the profile in Figure~\ref{fig:gasradprof}b.   Since $\rgas<r_{25}$, each galaxy includes $<$20 simulated ULXs (specifically from 5--18 per galaxy).   {\tt MARX} simulations are then run for each ULX, assuming the above positions on the ACIS-S3 chip, a point-source spatial model, and a spectral model consisting of a $\Gamma=1.8$ power law (including Galactic absorption) normalized to $L_X=10^{39}~\ergs$ (0.3-10~keV).   The simulations for the diffuse emission and each ULX  are then combined (using {\tt marxcat}), and corresponding simulated event and aspect solution files are created with {\tt marx2fits} and {\tt marxasp}, respectively.    We  then generate 0.3-7~keV band images from the simulated event files for each galaxy.
 Finally, we run {\tt wavdetect} on the simulated images with the same parameters as in \S\ref{sec:data}.   We simulate a total 214 ULXs combining all 20  galaxies, all of which are recovered by running {\tt wavdetect} on the simulated 0.3-7~keV images.\footnote{We found that running {\tt wavdetect} with wavelet scales larger than 2.0 pixels led to false identifications of a small number  of extended gas clumps as ULX candidates ($<$5), which is why we run {\tt wavdetect} with scales of 1.0, 1.4, and 2.0 pixels when assembling the ULX catalog.} 
  Thus, we conclude from these simulations that the diffuse gas emission is not strong enough in any of the AMUSE galaxies to dilute emission from a ULX to the extent that {\tt wavdetect} could no longer identify the X-ray source. 
      
 After identifying X-ray sources from the 0.3-7~keV band images, ULX candidates embedded in diffuse emission are ultimately classified via photometry in 2-7~keV hard-band images (see \S \ref{sec:data}).   We  note that the relatively hotter gas component for VCC1316 (M87; $kT_g=1.6\pm0.02$~keV) contributes substantially more photons to the 2--7~keV band compared to the other galaxies (we estimate that the gas component could contribute up to 30\% of the total flux contained within a typical photometric  aperture for a 10$^{39}~\ergs$ ULX candidate near the center of VCC1316, while the contribution is smaller for ULXs that are farther from the center and/or more luminous).  However, our local background subtraction in the 2-7~keV images adequately removes the gas component from each point source in VCC1316, and no other galaxy has gas that is hot enough to contribute so strongly to the hard X-rays.  However, for all galaxies where we use hard-band images, we must make sure that ULXs  are  still  above our  detection limit.   For each of the 20 galaxies with gas emission, plus the three additional galaxies with crowded centers, we use PIMMS to estimate the  expected 2-7~keV count rate  from a $10^{39}~\ergs$ (0.3-10~keV) ULX, assuming the distance to each galaxy, Galactic absorption, and a $\Gamma=1.8$ powerlaw.   Given the exposure times of each \textit{Chandra} observation, we expect an obvious detection for a ULX in every hard-band image  (with at least seven hard X-ray counts; six counts if we assume $\Gamma=2.2$),  except for perhaps NGC5077 (obsID=11780; $\tau_{\rm exp}=28.5$~ks; $d=40.2$~Mpc).    For NGC5077 we would still expect four hard X-ray counts (three if $\Gamma=2.2$) from a $10^{39}~\ergs$ ULX, which would usually be considered a marginal detection depending on the background level.  Upon visual  inspection of the (non-simulated) data, we note that there are only two {\tt wavdetect} sources in NGC5077 near significant amounts of gas.  Both of these sources have five net counts in the hard-band and are considered detections based on their background level of $<$0.5 counts \citep{kraft91}. Thus, we are unlikely incomplete to any ULXs due to our choice of using the hard X-ray band to perform photometry in the inner regions of these 23 galaxies.

 \section{Results}
 \label{sec:results}

 \subsection{Cluster vs.\ Field ULX Number Counts and Specific Frequencies}
 \label{sec:numcounts}
For each galaxy, we statistically correct for the expected background contamination and for the (small) number of ULXs that might not be covered by the S3 chip, and we calculate the net number of ULXs as $n^{i}_{\rm ulx} = \left(N^i_{\rm ulx} - N^i_{\rm bg}\right)/f^i_{\rm ulx}$, where $i$ refers to the $i^{\rm th}$ galaxy.  We calculate a total of $n_{\rm ulx} = \sum\nolimits_i n^{i}_{\rm ulx} =  37.9\pm 10.1$ and   $28.1\pm  8.7$ ULXs in the 99 Virgo and 96 Field galaxies, respectively.  (If we do not require $0\leq N^i_{\rm bg} \leq N^i_{\rm ulx}$, then we find $n_{\rm ulx} = 33.6\pm  9.5$  and $20.3\pm   7.4$ in Virgo and the Field, respectively).  However, care must be taken when comparing these ULX number counts, since each sample has different stellar mass distributions in addition to  a different number of galaxies (see Figure~1).  We control for the different $\mstar$ distributions following the method developed by \citet[][see their \S2.2]{miller12}.   From $\log \mstar$ histograms of each sample, we use the ratio of the number of Virgo to Field galaxies to weight the Field distribution to match that of Virgo (see Figure~2c in \citealt{miller12}).  We then randomly draw 10$^3$ subsamples from the weighted Field $\mstar$ distribution, with each subsample consisting of 45 galaxies.\footnote{As described by \citet{miller12}, $n=45$ is a practical limit to the subsample size because each subsample is drawn without replacement.} 
 In order to compare a similar number of Virgo and Field galaxies, we also create $10^3$ Virgo subsamples by randomly drawing 45 galaxies from Virgo each time (with no weighting).  The 10$^3$ Virgo and 10$^3$ Field subsamples have median net ULX counts of $n_{\rm ulx} = 7.2^{+4.8}_{-4.3}$ and $8.6^{+5.8}_{-4.3}$, respectively, where the error is the 90\% confidence interval based on the $n_{\rm ulx}$ distributions of all 10$^3$ subsamples.   We also repeat this exercise dropping the requirement that $0 \leq N_{\rm bg} \leq N_{\rm ulx}$, and we find median mass-matched net ULX counts of $n_{\rm ulx} = 5.1^{+5.2}_{-4.6}$ and $5.0^{+6.4}_{-4.8}$ for Virgo and the Field, respectively.  We thus do not see any statistical evidence for a significant difference between the ULX count rates in Virgo and in the Field.  
 
\begin{figure*}
\begin{center}
\includegraphics[scale=0.95]{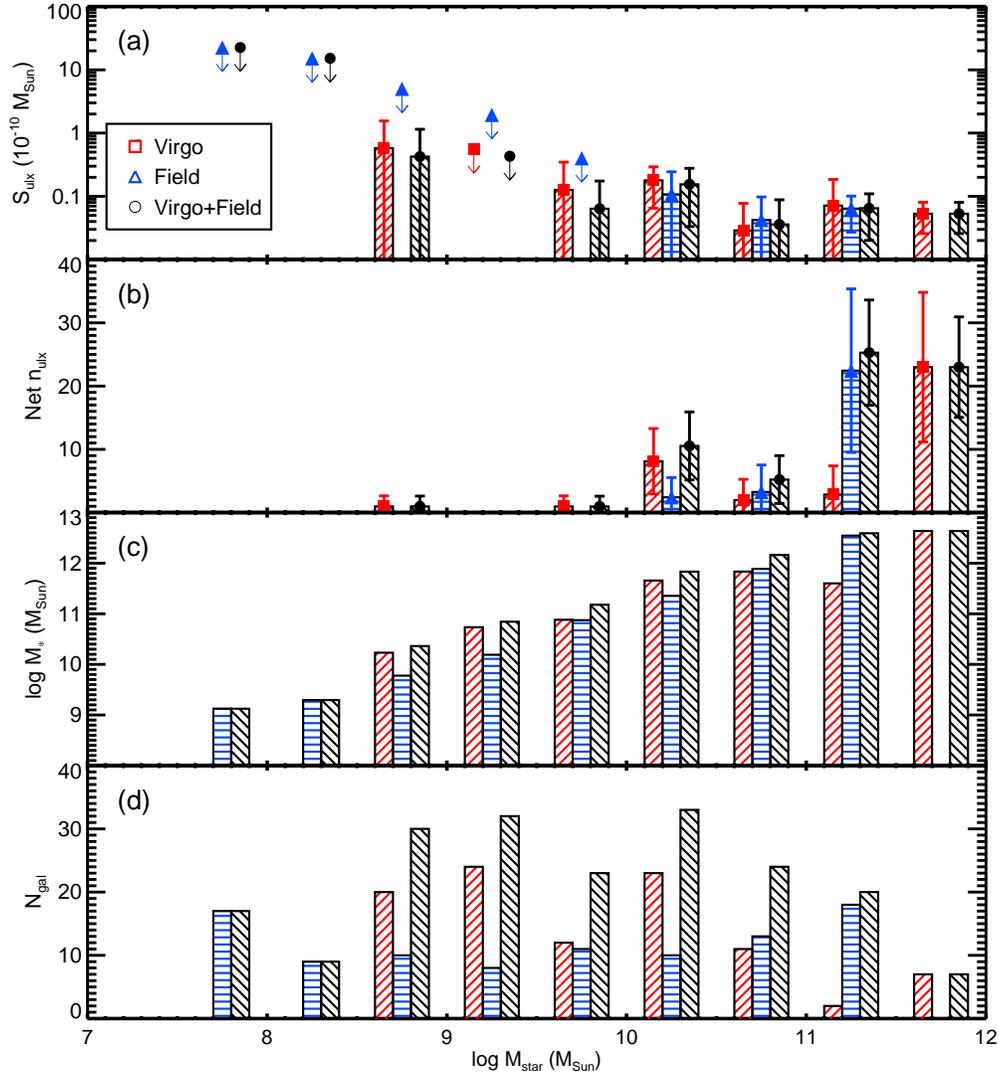}
\caption{(a) Histograms of the specific ULX frequency versus $\log \mstar$ for Virgo (red squares), the Field (blue triangles), and  the combined Virgo+Field AMUSE sample (back circles).  The three histograms within each mass bin are offset along the x-axis for clarity (plotted as Virgo, Field, and Virgo+Field from left to right).  Symbols at the top of shaded histograms mark mass bins containing ULXs (with 90\% confidence error bars), while the other symbols mark upper limits (95\% confidence).   Histograms are omitted for mass bins that do not contain any ULXs and/or galaxies.   (b) The net number of ULXs ($n_{\rm ulx}$) in each mass bin.  (c) The total amount of stellar mass contained in each mass bin.  (d) The total number of galaxies ($N_{\rm gal}$) in each mass bin.  }
\label{fig:ulxrates}
\end{center}
\end{figure*}

In addition to ULX number counts, it is  insightful to consider the specific ULX frequency, which we define as $S_{\rm ulx}=n_{\rm ulx}/\mstar^{10}$, where $\mstar^{10}$ is the stellar mass in units of $10^{10}~\msun$.    In Figure~\ref{fig:ulxrates}, we compare the specific ULX frequencies in each sample over different mass bins. For mass bins without any ULXs, we estimate 95\% confidence upper limits to $S_{\rm ulx}$ based on the total stellar mass contained within each bin.   We  test for the presence of a correlation between  $\mstar$ and $S_{\rm ulx}$ using the survival analysis package ASURV\footnote{\url{http://astrostatistics.psu.edu/statcodes/asurv}} 
Rev 1.2 \citep{lavalley92}, which implements the methods presented in \citet{isobe86}.   There is no statistically significant correlation in either sample ($p=0.08$ and 0.60 that no correlation is present in Virgo and the Field, respectively, from the generalized Kendall's $\tau$ test).   For completeness, we also include  the combined Virgo+Field sample (195 galaxies) in Figure~\ref{fig:ulxrates}, for which there is also no statistically significant correlation between $S_{\rm ulx}$ and $\mstar$  ($p=0.19$).   The lack of ULXs in the largest mass bin for the Field is because the Field sample does not include any galaxies with $\log\mstar>11.5~\msun$.  The detection of only two ULXs  (across both samples) hosted by galaxies with $\mstar<10^{10}~\msun$ could indicate that ULXs are extremely rare in lower-mass early-type galaxies.  However, we  alternatively may simply not be probing enough low-mass galaxies to detect many ULXs  (each mass bin contains $<10^{11}~\msun$ total, and we expect on the order of only one ULX per $\sim$$10^{11}~\msun$ in ellipticals; \citealt{walton11}), which we will discuss in \S \ref{sec:discdwarf}. 

  Finally, we note that combining all of the mass bins in each AMUSE sample (amounting to total stellar masses of  $6.0\times10^{12}$ and $4.6\times10^{12}~\msun$ in Virgo and the Field, respectively), the Virgo and Field samples  have nearly identical specific ULX frequencies of $S_{\rm ulx}=0.063 \pm  0.017$ and $0.061 \pm 0.019$~per $10^{10}~\msun$, further indicating that the ULX population in early galaxies does not depend on environment.  The specific frequency for all 195 galaxies is $S_{\rm ulx} = 0.062\pm0.013$ per 10$^{10}~\msun$ (from the literature, we expect on the order of $S_{\rm ulx}\approx0.1$; \citealt{walton11}).\footnote{If we do not require $0 \leq N_{\rm bg} \leq N_{\rm ulx}$, then the specific ULX frequencies are lower: $0.056 \pm 0.016$, $0.044 \pm 0.016$, and $0.051 \pm 0.011$ per 10$^{10}~\msun$ for Virgo, the Field, and Virgo+Field, respectively.}  
    We  note that we are potentially systematically underestimating $S_{\rm ulx}$ since we only count ULXs within the $D_{25}$~isophote, but the stellar masses are integrated over the entire galaxy.  However, calculating  $S_{\rm ulx}$ in this manner allows a more uniform comparison to the bulk of the literature, and we expect any potential bias to be negligible since only a very small number of ULXs likely fall outside of the $D_{25}$ isophote.    We also note that these specific ULX frequencies are not highly sensitive to the different $\mstar$ distributions of each sample, since we do not see a trend between $S_{\rm ulx}$ and $\mstar$ in Figure~\ref{fig:ulxrates}.

 \subsection{Average Spectral Properties}
 A detailed study on the spectral properties of ULXs is not possible, since most ULX candidates in the AMUSE survey have $<$$10^2$ X-ray counts.   Instead, we perform joint spectral fits to determine if there is a difference between the average spectral properties of ULXs in the Virgo versus the Field samples.   We note that other studies on individual ULXs utilizing higher quality X-ray spectra  indicate  diverse spectral properties that could reflect differences in, e.g., accretion geometries, outflow strengths, viewing angles, etc. \citep[see, e.g., \S 4 of][and references therein]{feng11}.   The heterogeneous nature of ULX spectra thus imposes a systematic limitation to the quality of our joint fits, which must be kept in mind when interpreting the results.    We thus only attempt to constrain the ``average'' ULX spectral properties at a phenomenological level here.
 
 For each ULX candidate in Tables~\ref{tab:virgoulx}-\ref{tab:Fieldulx}, we extract an X-ray spectrum with the {\tt CIAO} task {\tt specextract}.   Each source's spectrum is extracted within a circular region 2 pixels larger in radius than an aperture containing the 90\% encircled energy fraction.   We extract a local background from an annulus with inner and outer radii 5 and 15 pixels larger than the source extraction region.  For a handful of cases in crowded fields, we manually adjust the extraction regions to avoid contamination from nearby sources.   We generate (unweighted) arfs and rmfs for each spectrum, applying an energy dependent point-source aperture correction based on the size of the extraction region and the position on the S3 chip.   We simultaneously fit a multiple blackbody accretion disk model ({\tt phabs*diskbb}) to all ULX spectra in Virgo, and also to all ULX spectra in the Field, using Cash statistics (including the background in the fit),  allowing the normalization and intrinsic absorption to vary for each spectrum.\footnote{During this process, we found that six ULX candidates in Virgo and  five in the Field include significant amounts of gas within their spectral extraction regions.  Unfortunately, the total number of counts for these sources is too low to  include a gas model to the fits, and we similarly cannot reliably control for this gas component from the local background over the full 0.3-7.0~keV band (although we note that the local background is reliably controlled for when performing photometry over the hard 2-7~keV band, so that their identifications as ULXs are secure).  Thus, we exclude these contaminated spectra from the joint spectral fits, and we also exclude the four Virgo and one Field sources identified as stars/AGN in \S\ref{sec:bg}.  We thus fit a total of 45 and 44 spectra for Virgo and the Field, respectively.} 
 We find inner disk temperatures of $kT_{\rm in}=1.28\pm 0.07$~keV (Cash statistic $C=1372/1103$ degrees of freedom) and $kT_{\rm in} = 1.20^{+0.06}_{-0.02}$~keV ($C=1854/1047$ degrees of freedom) for Virgo and the Field, respectively.    We also simultaneously  fit an absorbed powerlaw model ({\tt phabs*powerlaw}) to each subsample, again allowing the intrinsic absorption and normalization to vary for each spectrum.  We obtain best-fit  photon indices of $1.67\pm0.07$  ($C=1120/1103$ degrees of freedom) for Virgo and $1.77\pm0.04$ ($C=1210/1047$ degrees of freedom)  for the Field.   Since both sets of inner disk temperatures and photon indices are similar,  we do not have evidence that the typical ULX spectrum in early-type galaxies is substantially different in various large-scale environments.   
   
 \section{Discussion}
 \label{sec:disc}
 
  The above results are consistent with  ULXs in early-type galaxies representing the luminous tail of the LMXB population, with little to no dependence on environment.  All of our ULX candidates have $L_X < 1.3\times10^{40}~\ergs$ ($<8.5\times10^{39}~\ergs$~excluding the five obvious contaminants identified in \S\ref{sec:bg}), and we therefore do not find any exotic IMBH candidates with $L_X > 10^{41}~\ergs$ within the AMUSE survey.    Furthermore, consistent with \citet{irwin04}, we find no statistical evidence for a population ULXs with $L_X > 2\times10^{39}~\ergs$ in early-type galaxies.   Our sample includes only 13 (Virgo) and 12 (Field) ULXs with $L_X > 2\times10^{39}~\ergs$, all of which could be  attributed to being unrelated foreground/background sources (statistically, we expect a total of 17.6 and 21.9 contaminants in Virgo and the Field, respectively).  Indeed, these 25 sources include the 5  sources already identified as a star or AGN from SIMBAD/NED.  The paucity of  very luminous sources  in early-type galaxies is understood if luminosities in excess of a few times 10$^{39}~\ergs$ require mass transfer from a high-mass companion star \citep[e.g.,][]{king02}.  These types of sources should therefore be short-lived, and associated with HMXBs in high-SFR environments (HMXBs tend to have ages of only $\sim$50~Myr; see \citealt{williams13} and references therein). 
  
Our ULX candidates are likely LMXBs observed toward the peak of an X-ray outburst, when the LMXB is in the very high state (see, e.g., \citealt{remillard06, fender04, fender09} for reviews on LMXB outburst phenomenology).    We thus expect relatively soft spectra.  Indeed, our best-fit disk temperatures of 1--2~keV are typical for soft-state LMXBs, and from the $M \propto T^{-4}$ relation would seem to exclude sub-Eddington IMBHs.\footnote{If stellar mass black holes,  we cannot unambiguously estimate black hole masses from the best-fit disk temperatures.  Such high-Eddington ratio sources could instead be accreting from a geometrically thick `slim' disk \citep{abramowicz88} and would follow a relation flatter than $M\propto T^{-4}$.  Also see, e.g., \citet{miller13} for other  caveats.} 
The average best-fit spectral photon indices ($\Gamma=1.67\pm0.07$ in Virgo and $\Gamma=1.77\pm0.04$ in the Field) are also consistent with the range of photon indices observed for other ULXs  in early-type galaxies that do not require invoking an IMBH to understand (e.g., \citealt{swartz04, berghea08, brassington10}).

Our joint spectral fits probably include some ULX candidates that are actually unidentified background AGN, which could potentially bias our best-fit photon indices.   We therefore repeat the joint spectral fits, and we attempt to reduce the background contamination by excluding all ULX candidates with $L_X > 2\times10^{39}~\ergs$ from the fits (note that  we cannot identify all contaminants on a case-by-case basis, but we expect that most  will be background AGN).  Also, since ULXs should be concentrated toward the center of the galaxy and background sources will be distributed uniformly across the sky, we additionally exclude any ULX candidate that is located at a distance $r/r_{25}>\sqrt{0.5}$ from the center of the galaxy (which should remove around half of the remaining contaminants).  These cuts remove 16 Virgo and 21 Field sources, which is on the same order of the number of expected contaminants from the cosmic X-ray background.  After excluding additional sources embedded in too much gas to reliably account for their local background, we refit both the absorbed disk and absorbed powerlaw models to a total of 36 Virgo and 26 Field ULX candidates.   We find similar results as before, with $kT_{\rm in}=1.29\pm 0.07$~keV  ($C=1130/902$ degrees of freedom) for Virgo and $kT_{\rm in}=1.22\pm 0.06$~keV ($C=878/557$ degrees of freedom)  for the Field, and $\Gamma=1.64\pm0.08$  ($C=926/902$ degrees of freedom) for Virgo and $\Gamma=1.78\pm0.08$ ($C=642/557$ degrees of freedom)  for the Field.  Thus, our spectral fits are unlikely highly biased by background AGN.   
  
 \subsection{ULXs in dwarf ellipticals}
 \label{sec:discdwarf}

There is currently no compelling evidence in the literature for a significant population of ULXs in dwarf elliptical galaxies, which could be due to previous studies not searching through enough early-type galaxies with low stellar-mass.       Following \citet{swartz08}, we define any galaxy with a stellar mass $\mstar<10^{9.3}~\msun$ as a dwarf (opposed to the more traditional absolute magnitude threshold $M_B > -18$), in order to  ease comparisons with the ULX literature.  For reference, two of the largest recent ULX surveys --- \citet{walton11} with \textit{XMM-Newton} and \citet{liu11} with \textit{Chandra} --- search through 97 and 65 elliptical galaxies respectively, and we estimate that $<$15\% of those are dwarfs (based on  the \citealt{bell03} relations and  $B-V$ colors  taken from a combination of their published catalogs and \citealt{de-vaucouleurs91}).\footnote{None of the 165 unique ULX detections hosted by ellipticals in the \citealt{walton11} catalog appears to  live in a galaxy with $\mstar < 10^{9.3}~\msun$.   There is one ULX candidate in the \cite{liu11} catalog with $L_X=1.2\times10^{40}$ -- CXOJ100633.962-295617.02 hosted by NGC3125, which we estimate to have $\mstar=10^9~\msun$.  Although NGC3125 is classified as an elliptical galaxy by \citet{liu11}, as taken from \citet{de-vaucouleurs91}, it is actually a (starbursting) dwarf irregular \citep[see, e.g.,][]{hadfield06}.} 
Various sample selection effects also limit the number of dwarf elliptical galaxies included in the large \citet{swartz08, swartz11} ULX surveys to even smaller numbers.  

Combining  the Virgo and  Field AMUSE samples, our study includes 72 dwarf ellipticals containing a total stellar mass of $\mstar=5\times10^{10}~\msun$.  This total mass in dwarf ellipticals represents almost an order of magnitude improvement in stellar mass over the \citet{walton11} survey, as estimated from their Figure~10.   We find only one ULX candidate in a dwarf elliptical galaxy in Virgo (which is  unlikely a background/foreground source) and none in the Field.   Thus, even with our larger sample we do not unambiguously identify a population of  ULXs in dwarf ellipticals.  Plus, our lone ULX candidate is hosted by VCC1779 ($\mstar=10^9~\msun$), which \citet{ferrarese06} note has clumpy dust in an \textit{HST} image.  \citet{ferrarese06} speculate that VCC1779 may actually be a dwarf irregular in the process of transitioning into a dwarf elliptical.   If we exclude VCC1779 from our sample, then we would not have recovered any ULXs in dwarf ellipticals.  Whether or not we include this galaxy does not qualitatively change our conclusions, since even one ULX is still an insignificant population.    Since we find that elliptical galaxies have $S_{\rm ulx}=0.062$ per 10$^{10}~\msun$, meaning we expect about 1 ULX per $1.6\times10^{11}~\msun$, we would need to survey at least three times more dwarf ellipticals to have a realistic chance of finding their  ULX population.

Although the existence of ULXs in dwarf ellipticals remains unclear, we do search through enough stellar mass to determine that the ULX populations in dwarf ellipticals and in dwarf spirals are statistically different.    Combining the Virgo and Field samples,  we find $S_{\rm ulx}=0.20^{+0.74}_{-0.19}$ per 10$^{10}~\msun$ in dwarf ellipticals if we include VCC1779, and $S_{\rm ulx}<0.61$ per 10$^{10}~\msun$ if we exclude VCC1779 (95\% confidence limit).  The exact ULX specific frequency in dwarf spirals is somewhat dependent on the particular sample, but dwarf spirals have a larger $S_{\rm ulx}\approx $1--5 per 10$^{10}~\msun$ (see Figure~2 in \citealt{swartz08} and Figure~10 in \citealt{walton11}).  Furthermore, when  also considering our improved constraints from $10^{9.3} < \mstar < 10^{10}~\msun$, the lack of any  correlation between $S_{\rm ulx}$ and $\mstar$ for elliptical galaxies (see \S \ref{sec:numcounts} and Figure~\ref{fig:ulxrates}) further indicates that the ULX population is different between dwarf ellipticals and dwarf spirals.\footnote{If we exclude VCC1779 from the survival analysis in \S\ref{sec:numcounts}, the statistical significance of any correlation between $S_{\rm ulx}$ and $\mstar$ becomes even less likely, with $p=0.33$ for Virgo (98 galaxies) and $p=0.62$ when combining both Virgo and the Field (194 galaxies).} 

  There is an active debate on whether dwarf ellipticals are the faint extension of the giant elliptical population \citep[e.g.,][]{graham03, ferrarese06}, or if there is a dichotomy between the two populations \citep[e.g.,][and references therein]{kormendy09}.   In the latter case, dwarf ellipticals may be morphologically, structurally, and kinematically more similar to late-type galaxies, potentially leading to a sequence where (some) dwarf spirals  evolve into dwarf ellipticals through a variety of  different  environmental mechanisms  \citep[e.g.,][]{faber83, moore98, boselli08}. The flat trend between $S_{\rm ulx}$ and $\mstar$ for all early-type galaxies, combined with the difference in $S_{\rm ulx}$ between dwarf ellipticals and dwarf spirals, may at first seem to indicate that dwarf and giant ellipticals form a continuous population.  However, a more likely scenario is  that  any ULX in a dwarf elliptical does not possess a `memory' of the galaxy's evolutionary history.   ULXs in dwarf ellipticals  probe  only an older stellar population, while dwarf spirals are undergoing enough current star formation for ULXs to also probe the shorter-lived HMXBs.   Even if dwarf spirals transition into dwarf ellipticals, after star formation is quenched, it is then  the total stellar mass that  primarily dictates how many LMXBs will radiate above 10$^{39}~\ergs$.  Indeed, the star formation in the AMUSE dwarf ellipticals is too small for HMXBs to contribute to their ULX population.  We check their SFRs by correlating the AMUSE galaxies to the All-Sky Data Release of the \textit{Wide-Field Infrared Survey Explorer} \citep[WISE;][]{wright10}, and we find matches (within 3$\arcsec$) to 41/72 of our dwarf ellipticals (accounting for a total $\mstar=3.3\times10^{10}~\msun$).   Their  WISE colors are consistent with the expected colors of other quiescent galaxies in Figure~12 of \citet{wright10}.  To convert the WISE colors to more physical units, we use the relation between specific star formation rate and WISE 4.6-12$\mu$m color in Equation 5 of  \citet{donoso12}.  No AMUSE dwarf elliptical shows a SFR larger than 0.021 $\msun$~yr$^{-1}$, and most  are not detected at 12$\mu$m yielding upper limits on their SFRs (typically with SFR$<10^{-3}$~$\msun$~yr$^{-1}$).\footnote{We stress that the exact SFR values are uncertain because they are extrapolated from a relation calibrated to massive starburst galaxies, and we quote them simply as a means to report SFRs in more familiar units than WISE color, and to illustrate that any star formation is relatively weak.}    
  
 \section{Summary}
We perform the largest  study to date on the ULX population in early-type galaxies, making use of the homogeneous X-ray coverage of galaxies included in the AMUSE survey. In particular, we focus on whether the properties of ULXs in elliptical galaxies depend on galactic environment.  Searching through 99 galaxies in the Virgo cluster and 96 galaxies in the field, we respectively identify a total of 55 and 50 non-nuclear X-ray point sources with $L_X > 10^{39}~\ergs$ (0.3-10~keV).   Accounting for contamination from the cosmic X-ray background, we calculate nearly identical specific ULX frequencies of $S_{\rm ulx} = 0.063\pm 0.017$ and $0.061\pm 0.019$ ULXs per 10$^{10}~\msun$ in the Virgo and Field samples, respectively.   We find that there are a similar number of ULXs in each environment, after correcting for the different stellar mass distributions of each sample.   The average X-ray spectral shapes of the ULX candidates in each sample are also similar.  We thus find no evidence for an environmental dependence on the ULX population in early-type galaxies.  

Our results are consistent with ULXs in early-type galaxies composing the high-luminosity tail of the galaxy's LMXB population, meaning that the total number of ULXs in an early-type galaxy should scale primarily with the galaxy's stellar mass \citep{gilfanov04}.   Support for this conclusion includes the lack of any correlation between specific ULX frequency and stellar mass, and we also do not find a meaningful population of ULXs with $L_X>2\times10^{39}~\ergs$ \citep{irwin04}.   Combining both the Virgo and Field samples, we calculate $S_{\rm ulx}=0.062\pm 0.013$ per 10$^{10}~\msun$, or on the order of one ULX per $\sim$$1.6\times10^{11}~\msun$, which is similar to  previous constraints on ULXs in early-type galaxies  \citep[e.g.,][]{walton11}.   We do not probe enough stellar mass in the lowest-$\mstar$ galaxies to determine if dwarf ellipticals  host ULXs.  However, we do place the tightest constraints on the ULX population in dwarf ellipticals so far ($S_{\rm ulx}<0.61$ per 10$^{10}~\msun$), and we determine that they must contain  fewer ULXs per unit stellar mass than dwarf spiral galaxies.

ULXs in early-type galaxies are likely composed of a less heterogeneous population of sources than ULXs in late-type galaxies.   Thus, with relatively modest X-ray exposure times, ULXs in ellipticals offer a clean probe of a galaxy's LMXB population, and a means to determine what galaxy properties beyond star formation rate may influence stellar mass black hole production and subsequent activity.  We conclude that environment does not play a very strong role on ULX rates, and that stellar mass is the most important factor.  Metallicity is potentially another important factor, and the fraction of black holes  with low-mass binary companions is also important  for determining the number of actively accreting stellar mass black holes in a galaxy.   In the future, X-ray observations of even more dwarf elliptical galaxies are needed to determine if they  host ULXs.   Studies comparing ULXs to other early-type galaxy properties like, e.g., metallicity or globular cluster frequency  would also be illuminating.

\acknowledgments
We thank the anonymous referee for constructive feedback that improved this manuscript, and we thank Michael Katolik for contributing to our selection algorithms.  Support for this work was provided by the National Aeronautics and Space Administration (NASA) through Chandra Award Number 11620915 issued by the Chandra X-ray Observatory Center, which is operated by the Smithsonian Astrophysical Observatory for and on behalf of  NASA under contract NAS8-03060.  Support was also provided by NASA through a grant from the Space Telescope Science Institute associated with program HST-GO-12591.01-A, which is operated by the Association of Universities for Research in Astronomy, Inc., under NASA contract NAS 5-26555.  T.T. acknowledges support from the Packard Foundation through a Packard Research Fellowship.  J.H.W. acknowledges  support by the National Research Foundation of Korea (NRF) grant funded by the Korea government (MEST; No. 2012-006087).   This research has made use of data obtained from the Chandra Data Archive and the Chandra Source Catalog, and software provided by the Chandra X-Ray Center (CXC) in the application packages CIAO, ChIPS, and Sherpa.  This publication makes use of data products from the Wide-field Infrared Survey Explorer, which is a joint project of the University of California, Los Angeles, and the Jet Propulsion Laboratory/California Institute of Technology, funded by NASA.



\clearpage
\LongTables
\renewcommand\arraystretch{1.1}
\begin{deluxetable*}{l r@{.}l r@{.}l  c c c c c c c c c  c c c}
\tablecaption{Properties and ULX count rates for 99 AMUSE-Virgo galaxies \label{tab:virgoprop}}
\tablewidth{18cm}
\tabletypesize{\scriptsize}

\tablehead{
                \colhead{Galaxy}      		&  
                \multicolumn{2}{c}{R.A.}      &  
                \multicolumn{2}{c}{Dec.}         		&  
                \colhead{d}             &   
                \colhead{$\log \mstar$}                      &  
                \colhead{\textit{Chandra}}    & 
                \colhead{$r_{25}$}		&  
                \colhead{$e$} 	&
                \colhead{P.A.}                & 
                \colhead{$f_{\rm area}$}   &  
                \colhead{$f_{\rm ulx}$} & 
                \colhead{$N_{\rm bg}$} &  
                \colhead{$N_{\rm ulx}$} & 
                \colhead{$n_{\rm ulx}$} \\ 
	      \colhead{Name}	 			&  
	      \multicolumn{2}{c}{(deg.)}			& 
	      \multicolumn{2}{c}{(deg.)}			& 
	      \colhead{(Mpc)}		& 
	      \colhead{($\msun$)}	 & 
	      \colhead{ObsID}           & 
	      \colhead{($\arcmin$)}           &  
	      \colhead{}              & 
	       \colhead{(deg)}    & 
	       \colhead{}	                 & 
		\colhead{}                 	& 
		\colhead{}	         & 
		\colhead{}	         & 
		\colhead{}      \\ 
		\colhead{(1)} &
		\multicolumn{2}{c}{(2)} &
		\multicolumn{2}{c}{(3)} &
		\colhead{(4)} &
		\colhead{(5)} &
		\colhead{(6)} &
		\colhead{(7)} &
		\colhead{(8)} &
		\colhead{(9)} &
		\colhead{(10)} &
		\colhead{(11)} &
		\colhead{(12)} & 
		\colhead{(13)} &
		\colhead{(14)}  \\
}

\startdata
                  VCC1226 &    $187$ &   444946 &      $8$ &   000417 &    17.14 &     12.0 &      321 &     5.24 &     0.58 &      157 &    0.725 &    0.951 &     2.40 &        6 & $3.79^{+5.12}_{-3.30}$ \\ 
                   VCC881 &    $186$ &   548904 &     $12$ &   946222 &    16.83 &     11.9 &      318 &     6.01 &     0.69 &      127 &    0.660 &    0.922 &     2.44 &        3 & $0.61^{+4.45}_{-0.61}$ \\ 
                  VCC1316 &    $187$ &   705917 &     $12$ &   391111 &    17.22 &     11.8 &     3717 &     3.71 &     0.36 &        0 &    0.870 &    0.985 &     1.68 &        9 & $7.44^{+6.03}_{-4.17}$ \\ 
                  VCC1978 &    $190$ &   915253 &     $11$ &   552611 &    17.30 &     11.7 &      785 &     3.46 &     0.61 &      104 &    0.939 &    0.996 &     1.36 &        7 & $5.66^{+5.39}_{-3.54}$ \\ 
                   VCC763 &    $186$ &   265503 &     $12$ &   887000 &    18.45 &     11.7 &     6131 &     3.88 &     0.49 &      132 &    0.970 &    0.998 &     2.30 &        4 & $1.70^{+4.16}_{-1.70}$ \\ 
                   VCC731 &    $186$ &   117493 &      $7$ &   317500 &    23.33 &     11.7 &     5921 &     2.62 &     0.69 &       40 &    1.000 &    1.000 &     1.63 &        4 & $2.37^{+4.10}_{-2.31}$ \\ 
                   VCC798 &    $186$ &   350174 &     $18$ &   190527 &    17.86 &     11.6 &     2016 &     3.54 &     0.63 &        0 &    0.981 &    0.999 &     1.58 &        3 & $1.42^{+3.82}_{-1.42}$ \\ 
                  VCC1903 &    $190$ &   509995 &     $11$ &   646667 &    14.93 &     11.3 &     2068 &     2.34 &     0.71 &      164 &    1.000 &    1.000 &     0.39 &        1 & $0.61^{+2.98}_{-0.61}$ \\
\enddata

\tablecomments{\footnotesize Rows are sorted in order of decreasing stellar mass.  Columns 1--6 are taken directly from \citet{gallo08,gallo10} (also see \S \ref{sec:obs}). 
 Column 7: the radius of the semi-major axis of each galaxy's $D_{25}$ isophote in the $B$-band corrected for extinction, taken from the HyperLeda database.  
 Column 8: the eccentricity of the $D_{25}$ isophote, calculated from the ratio of semi-major to semi-minor axis listed in the HyperLeda database. 
 Column 9: position angle of the $D_{25}$ isophote's major axis, measured in degrees eastward of north.
Column 10: areal fraction of each galaxy's $D_{25}$ isophote that is covered by the ACIS-S3 chip (see \S \ref{sec:frac}).  
Column 11: fraction of ULXs within each $D_{25}$ isophote expected to be covered by the ACIS-S3 chip (see \S \ref{sec:frac}).  
Column 12: the number of background sources with $L_X > 10^{39}~\ergs$ statistically expected to fall within the fraction of each galaxy's $D_{25}$ isophote covered by the ACIS-S3 chip, based on the cosmic X-ray background from \citet{moretti03} (see \S \ref{sec:bg}).  
Column 13: the number of ULX candidates detected within the fraction of each $D_{25}$ isophote covered by the ACIS-S3 chip. 
 Column 14: the expected number of (net) ULXs expected for each galaxy, statistically corrected for  background contamination and chip coverage.  $n_{\rm ulx} = \left(N_{\rm ulx} - N_{\rm bg}\right)/f_{\rm ulx}$ (see \S \ref{sec:numcounts}).  Uncertainties are quoted at the 90\% confidence level from Poisson statistics, taking the expected number of background sources into account using the Bayesian formalism by \citet{kraft91}.}

\end{deluxetable*}
\renewcommand\arraystretch{1}

\clearpage
\LongTables
\begin{deluxetable*}{l r@{.}l r@{.}l c c c c c c c c c c c c}
\renewcommand\arraystretch{1.1}
\tablecaption{Properties and ULX count rates for 96 AMUSE-Field galaxies\label{tab:Fieldprop}}
\tablewidth{18.5cm}
\tabletypesize{\scriptsize}
\tablehead{
                \colhead{Galaxy}      		&  
                \multicolumn{2}{c}{R.A.}      &  
                \multicolumn{2}{c}{Dec.}         		&  
                \colhead{d}             &   
                \colhead{$\log \mstar$}                      &  
                \colhead{\textit{Chandra}}    & 
                \colhead{$r_{25}$}		&  
                \colhead{$e$} 	&
                \colhead{P.A.}                & 
                \colhead{$f_{\rm area}$}   &  
                \colhead{$f_{\rm ulx}$} & 
                \colhead{$N_{\rm bg}$} &  
                \colhead{$N_{\rm ulx}$} & 
                \colhead{$n_{\rm ulx}$} \\ 
	      \colhead{Name}	 			&  
	      \multicolumn{2}{c}{(deg.)}			& 
	      \multicolumn{2}{c}{(deg.)}			& 
	      \colhead{(Mpc)}		& 
	      \colhead{($\msun$)}	 & 
	      \colhead{ObsID}           & 
	      \colhead{($\arcmin$)}           &  
	      \colhead{}              & 
	       \colhead{(deg)}    & 
	       	\colhead{}                 & 
		\colhead{}                 	& 
		\colhead{}	         & 
		\colhead{}         &  
		\colhead{}	         \\ 
		\colhead{(1)} &
		\multicolumn{2}{c}{(2)} &
		\multicolumn{2}{c}{(3)} &
		\colhead{(4)} &
		\colhead{(5)} &
		\colhead{(6)} &
		\colhead{(7)} &
		\colhead{(8)} &
		\colhead{(9)} &
		\colhead{(10)} &
		\colhead{(11)} &
		\colhead{(12)} & 
		\colhead{(13)} &
		\colhead{(14)}  \\
}

\startdata
                  NGC1407 &     $55$ &   049583 &    $-18$ &   580278 &    23.20 &     11.5 &      791 &     3.15 &     0.41 &        0 &    0.928 &    0.994 &     2.71 &        7 & $4.31^{+5.28}_{-3.50}$ \\ 
                  NGC2768 &    $137$ &   906250 &     $60$ &   037222 &    23.10 &     11.5 &     9528 &     2.94 &     0.92 &       92 &    1.000 &    1.000 &     1.10 &        2 & $0.90^{+3.48}_{-0.90}$ \\ 
                  NGC4125 &    $182$ &   025000 &     $65$ &   174167 &    22.90 &     11.4 &     2071 &     3.01 &     0.63 &       95 &    0.934 &    0.995 &     1.00 &        1 & $0.00^{+3.29}_{-0.00}$ \\ 
                  NGC5846 &    $226$ &   622083 &      $1$ &   605556 &    26.20 &     11.4 &      788 &     2.34 &     0.30 &        0 &    1.000 &    1.000 &     2.23 &        3 & $0.77^{+4.02}_{-0.77}$ \\ 
                  NGC3923 &    $177$ &   757083 &    $-28$ &   806111 &    20.40 &     11.4 &     9507 &     3.88 &     0.75 &       48 &    1.000 &    1.000 &     2.33 &        4 & $1.67^{+4.16}_{-1.67}$ \\ 
                  NGC4697 &    $192$ &   149583 &     $-5$ &   800833 &    17.80 &     11.4 &     4730 &     3.62 &     0.81 &       83 &    1.000 &    1.000 &     1.27 &        2 & $0.73^{+3.54}_{-0.73}$ \\ 
                  NGC4494 &    $187$ &   850417 &     $25$ &   775278 &    20.90 &     11.4 &     2079 &     2.23 &     0.30 &        0 &    1.000 &    1.000 &     1.19 &        2 & $0.81^{+3.51}_{-0.81}$ \\ 
                   IC1459 &    $344$ &   294167 &    $-36$ &   462222 &    23.20 &     11.4 &     2196 &     2.34 &     0.71 &       43 &    1.000 &    1.000 &     1.25 &        3 & $1.75^{+3.75}_{-1.75}$ \\
\enddata

\tablecomments{\scriptsize Columns 1--6 are taken directly from \citet{miller12a}.  Otherwise, see notes to Table \ref{tab:virgoprop}. }

\end{deluxetable*}
\renewcommand\arraystretch{1}

\clearpage
\begin{deluxetable}{l r@{.}l r@{.}l c c c c c c c}
\tablecaption{ULX Candidates in the Virgo Sample \label{tab:virgoulx}}
\tablewidth{17.5cm}
\tabletypesize{\scriptsize}

\tablehead{ 
                \colhead{Host}      		&  
                \multicolumn{2}{c}{R.A.}      &  
                \multicolumn{2}{c}{Dec.}         		&  
                \colhead{$r/r_{25}$}             &   
                \colhead{Net Counts}                      &  
                \colhead{Exp}		&  
                \colhead{Count Rate} 	&
                \colhead{Band}                & 
                \colhead{$f_X \times 10^{15}$}   &  
                \colhead{$\log L_X$} \\ 
	      \colhead{Galaxy}	 			&  
	      \multicolumn{2}{c}{(deg.)}			& 
	      \multicolumn{2}{c}{(deg.)}			& 
	      \colhead{}		& 
	      \colhead{}	 & 
	      \colhead{(ks)}           &  
	      \colhead{(ks$^{-1}$)}              & 
	       \colhead{}    & 
	       	\colhead{($\ergs~{\rm cm}^{-2}$)}                 & 
		\colhead{($\ergs$)}                 	\\ 
		\colhead{(1)} &
		\multicolumn{2}{c}{(2)} &
		\multicolumn{2}{c}{(3)} &
		\colhead{(4)} &
		\colhead{(5)} &
		\colhead{(6)} &
		\colhead{(7)} &
		\colhead{(8)} &
		\colhead{(9)} &
		\colhead{(10)} \\
}
\startdata
                  VCC1226 &    $187$ &   454214  &      $7$ &   980932 &     0.25 &         $27.63 \pm 8.65$ &     33.5 &              $0.824 \pm 0.258$ &     hard &        33.43 &         39.1 \\ 
                 \nodata  &    $187$ &   441978  &      $7$ &   987602 &     0.16 &         $22.45 \pm 7.79$ &     33.7 &              $0.665 \pm 0.231$ &     hard &        27.00 &         39.0 \\ 
                 \nodata  &    $187$ &   445557  &      $7$ &   998682 &     0.02 &         $24.86 \pm 8.20$ &     32.4 &              $0.768 \pm 0.253$ &     hard &        31.17 &         39.0 \\ 
                 \nodata  &    $187$ &   426329  &      $8$ &   002192 &     0.25 &        $48.99 \pm 11.51$ &     33.8 &              $1.449 \pm 0.341$ &     hard &        58.81 &         39.3 \\ 
                 \nodata  &    $187$ &   461945  &      $8$ &   002741 &     0.24 &         $22.25 \pm 7.76$ &     33.5 &              $0.664 \pm 0.232$ &     hard &        26.94 &         39.0 \\ 
                 \nodata  &    $187$ &   423289  &      $8$ &   004014 &     0.29 &         $23.01 \pm 7.89$ &     33.8 &              $0.681 \pm 0.234$ &     hard &        27.64 &         39.0 \\ 
                  VCC881 &    $186$ &   548845  &     $12$ &   947018 &     0.01 &         $17.23 \pm 6.83$ &     18.6 &              $0.926 \pm 0.367$ &     hard &        37.64 &         39.1 \\ 
                 \nodata  &    $186$ &   539326  &     $12$ &   951517 &     0.11 &  $13.91^{+7.86}_{-5.52}$ &     18.8 &      $0.741^{+0.419}_{-0.294}$ &     hard &        30.11 &         39.0 \\
\enddata
\tablecomments{Columns 1--3: host galaxy, right ascension and declination repeated from Table~\ref{tab:virgoprop}.  
Column 4: angular distance of each ULX candidate from the center of the galaxy, reported as the normalized semi-major axis of an elliptical isophote that includes the ULX, assuming concentric isophotes with the same eccentricity and position angle as the $D_{25}$ isophote.  
Column 5: total net X-ray counts for each ULX candidate.
Column 6: effective exposure time at the position of each ULX candidate (calculated from exposure maps for each \textit{Chandra} observation).  
Column 7: net count rate (in counts per ksec) for each ULX candidate.
Column 8: \textit{full}: photometry in columns 5 and 7 are reported from 0.3-7~keV; \textit{hard}: photometry in columns 5 and 7 are reported from 2-7~keV (see \S \ref{sec:data}).
Column 9: 0.3-10~keV X-ray flux, corrected for Galactic absorption, calculated using PIMMS and assuming a powerlaw with $\Gamma=1.8$  (see \S \ref{sec:ulxcat}).
Column 10: logarithm of  0.3-10~keV (unabsorbed) X-ray luminosity, from the fluxes in column 9 and the distances in Table~\ref{tab:virgoprop} (see \S \ref{sec:ulxcat}).
$^{\rm a}$Background AGN in NED/SIMBAD (\S \ref{sec:bg}).   $^{\rm b}$Foreground star SIMBAD (\S \ref{sec:bg}).  }

\end{deluxetable}

\clearpage
\begin{deluxetable}{l r@{.}l r@{.}l c c c c c c c}
\tablecaption{ULX Candidates in Field Sample \label{tab:Fieldulx}}
\tablewidth{17.5cm}
\tabletypesize{\scriptsize}

\tablehead{ 
                \colhead{Host}      		&  
                \multicolumn{2}{c}{R.A.}      &  
                \multicolumn{2}{c}{Dec.}         		&  
                \colhead{$r/r_{25}$}             &   
                \colhead{Net Counts}                      &  
                \colhead{Exp}		&  
                \colhead{Count Rate} 	&
                \colhead{Band}                & 
                \colhead{$f_X \times 10^{15}$}   &  
                \colhead{$\log L_X$} \\ 
	     \colhead{Galaxy} 	 			&  
	      \multicolumn{2}{c}{(deg.)}			& 
	      \multicolumn{2}{c}{(deg.)}			& 
	      \colhead{}		& 
	      \colhead{}	 & 
	      \colhead{(ks)}           &  
	      \colhead{(ks$^{-1}$)}              & 
	       \colhead{}    & 
	       	\colhead{($\ergs~{\rm cm}^{-2}$)}                 & 
		\colhead{($\ergs$)}                 	\\ 
}

\startdata
                  NGC1407 &     $55$ &   060510  &    $-18$ &   610503 &     0.61 &       $160.40 \pm 20.83$ &     40.8 &              $3.932 \pm 0.511$ &     full &        27.01 &         39.2 \\ 
                 \nodata  &     $55$ &   076108  &    $-18$ &   600480 &     0.65 &       $134.01 \pm 19.04$ &     42.6 &              $3.143 \pm 0.447$ &     full &        21.61 &         39.1 \\ 
                 \nodata  &     $55$ &   044233  &    $-18$ &   595323 &     0.31 &         $27.05 \pm 8.56$ &     43.0 &              $0.628 \pm 0.199$ &     hard &        25.67 &         39.2 \\ 
                 \nodata  &     $55$ &   052155  &    $-18$ &   581007 &     0.05 &        $41.62 \pm 10.61$ &     43.3 &              $0.961 \pm 0.245$ &     hard &        39.24 &         39.4 \\ 
                 \nodata  &     $55$ &   037162  &    $-18$ &   579938 &     0.25 &         $18.49 \pm 7.07$ &     41.1 &              $0.449 \pm 0.172$ &     hard &        18.35 &         39.1 \\ 
                 \nodata  &     $55$ &   038563  &    $-18$ &   578147 &     0.22 &         $17.13 \pm 6.81$ &     41.4 &              $0.413 \pm 0.164$ &     hard &        16.89 &         39.0 \\ 
                 \nodata  &     $55$ &   046785  &    $-18$ &   576389 &     0.09 &  $14.30^{+7.95}_{-5.60}$ &     41.6 &      $0.344^{+0.191}_{-0.135}$ &     hard &        14.05 &         39.0 \\ 
                  NGC2768 &    $137$ &   940141  &     $60$ &   026807 &     0.60 &       $110.43 \pm 17.29$ &     47.4 &              $2.329 \pm 0.365$ &     full &        19.09 &         39.1 \\
\enddata

\tablecomments{See Notes to Table \ref{tab:virgoulx}.   $^{\rm a}$Background AGN in NED/SIMBAD (\S \ref{sec:bg}). }

\end{deluxetable}

\end{document}